\documentclass[11pt]{article}  %Do not change the point size.
\usepackage{menuproc}
\usepackage{epsfig}
\usepackage{amsmath,amssymb}
\begin{document}
%
%____________________________________________________________
%
%  Title, authors, institutions, and abstract
%----------------------------------------------------------------
%  Syntax:  \titlematter{title}{authors}{institutions}{abstract}
%----------------------------------------------------------------
%     If lines are too long, use linebreaks where convenient.
%     If all authors are from the same institution, omit raised letters.
%
\titlematter{Modern Dyson-Schwinger Equation Studies}
{M.B.~Hecht and C.D.~Roberts}
{Physics Division, Argonne National Laboratory\\
Argonne, IL 60439, USA}
{\\
The dichotomy of the pion as QCD's Goldstone mode and a bound state of massive
constituents is easily understood using the Dyson-Schwinger equations. That
provides the foundation for an efficacious phenomenology, which correlates the
pion's charge radius and electromagnetic form factor with its valence quark
distribution function; and simultaneously provides a Poincar\'e covariant
description of the nucleon, its form factors and, more recently, meson
photoproduction processes.  This well-constrained framework can also be used to
eliminate candidates for an extension of the Standard Model by providing the
relation between current-quark electric dipole moments and that of the
neutron.\\
---
A summary of two presentations, one by each author.}
%
%____________________________________________________________
%  Start article here:
%%%%%%%%%%%%%%%%%%%%%%%%%%%%%%%%%%%%%%%%%%%%%%%%%%%%%%%%%%%%%%%%%%%%%%%%%%%%%%%%%%
\section{Introduction}
A focus of contemporary studies in QCD is the development of an intuitive
understanding of the spectrum and interactions of hadrons in terms of QCD's
elementary excitations; i.e., quarks and gluons.   Progress can be made by
applying a single framework to the calculation of many observables. This
facilitates a verification of necessary model assumptions, and the
identification of robust correlations between a theory's keystones and hadron
properties.  Non-hadronic electroweak interactions provide the obvious test bed
for any such approach because the electroweak probes are well understood and
hence a given experiment yields immediate access to properties of the hadron
target. Thus constrained the framework can be used reliably to make predictions
for other phenomena, even those far removed from the domain on which it was
constrained; e.g., the properties of QCD at nonzero temperature and baryon
density.

Herein we supply a brief description of recent progress with the
Dyson-Schwinger equations (DSEs)~\cite{cdragw} in these applications.  A
familiar DSE is the gap equation that describes pairing and condensation in low
temperature superconductors; another is the Bethe-Salpeter equation, whose
solution provides the mass and ``wave function'' of a bound state in quantum
field theory.  These are subjects in which the nonperturbative character of
DSEs is paramount, to which we will return.  However, at their simplest, the
DSEs are a generating tool for perturbation theory: the weak coupling expansion
of any particular DSE yields all the well-known Feynman diagrams. This is of
immense help in studying QCD because it means there is little or no model
dependence in the ultraviolet behaviour of calculated quantities.  The key
model-dependence is limited to the infrared domain, a property that can be
exploited to probe the dynamics underlying confinement and dynamical chiral
symmetry breaking (DCSB), which are QCD's signature nonperturbative phenomena.
The recent successes and current challenges in this application are documented
in Refs.~\cite{revbasti,revreinhard}.

\section{Dynamical Chiral Symmetry Breaking}
\begin{figure}[t]
\parbox{.55\textwidth}{\epsfig{file=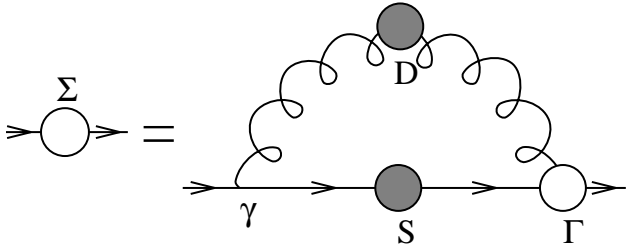,width=.5\textwidth,silent=,clip=}}
\hfill
\parbox{.45\textwidth}{\caption{\label{qcdgap} QCD Gap Equation or Dyson-Schwinger
equation for the quark self energy, $\Sigma$.  Here: $D$ is the dressed-gluon
propagator; $\Gamma$, the dressed-quark-gluon vertex; and $S=1/[i\gamma \cdot p
+ m + \Sigma(p)]$, the dressed-quark propagator.}}
\end{figure}

While the dynamical breaking of chiral symmetry in QCD is fundamental to the
success of chiral perturbation theory: it is only owing to DCSB that $m_\pi=0$
in the chiral limit and the scale of $f_\pi$ is set by the constituent-quark
mass, understanding the origin of this phenomenon is outside the scope of the
theory. This is where QCD's {\it gap equation}, depicted in Fig.~\ref{qcdgap},
finds immediate application~\cite{cdranu}.  The solution provides the
dressed-quark propagator in terms of the self-energy, $\Sigma$:
\begin{eqnarray}
S(p) &=& 1/[i\gamma\cdot p + m + \Sigma(p)]\,,\;\;
\Sigma(p) = i\gamma \cdot p \,[A(p^2)-1] + B(p^2)\,,
\end{eqnarray}
where $m$ is the current-quark mass, and the functions $A(p^2)$, $B(p^2)$ are
completely determined by the nature of the force between quarks.

%--------------------------------------------------------------------------------
%\begin{figure}[t]
%\centerline{\epsfig{file=qu_dseEucl.eps,width=0.50\textwidth,silent=,clip=}}
%\caption{\label{qcdgap} The QCD Gap Equation or Dyson-Schwinger equation for
%the quark self energy, $\Sigma$.  Here: $D$ is the dressed-gluon propagator;
%$\Gamma$, the dressed-quark-gluon vertex; and $S=1/[i\gamma \cdot p + m +
%\Sigma(p)]$, the dressed-quark propagator. }
%\end{figure}
%--------------------------------------------------------------------------------
%--------------------------------------------------------------------------------

Employing a weak coupling expansion yields the perturbative results:
$A(p^2)\approx 1$ and
\begin{equation}
B(p^2) = m \left( 1 - \frac{3\alpha_s}{4\pi}\ln[p^2/m^2] + \ldots \right)\,,
\end{equation}
where the ellipsis indicates that, like the zeroth and first order
contributions, every higher-order term in the perturbative evaluation of
$B(p^2)$ is proportional to the current-quark mass.  Hence at every finite
order in perturbation theory the scalar piece of the quark's self-energy,
$B(p^2)$, vanishes in the chiral limit, $m=0$.  Therefore dynamical mass
generation, and hence DCSB, is impossible in perturbation theory.

\begin{figure}[t]
\vspace*{-4ex}
\parbox{.55\textwidth}{\epsfig{file=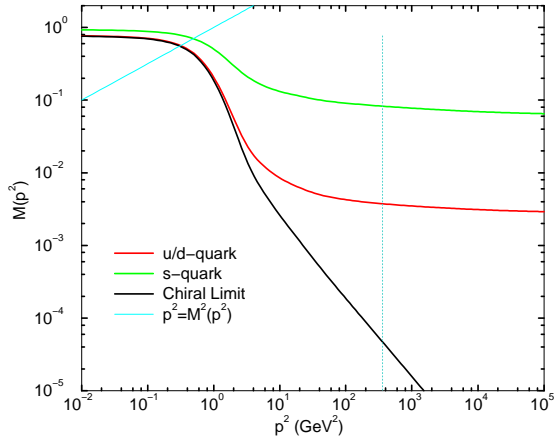,width=.5\textwidth,silent=,clip=}}
\hfill
\parbox{.45\textwidth}{\caption{\label{quarkM} Quark mass function: $M(p^2)=B(p^2)/A(p^2)$, in the
chiral limit, and for the $u$-quark (isospin symmetry is assumed):
$m_u^{1\,{\rm GeV}}=5.5\,$MeV, and the $s$-quark: $m_s^{1\,{\rm
GeV}}=130\,$MeV.  The intersection of the line $p^2=M^2(p^2)$ with a given
curve gives the Euclidean constituent-quark mass~\protect\cite{mr97}.}}
\end{figure}

Now the essentially nonperturbative character of DSEs becomes important.  The
self-energy appears in the denominator of the r.h.s.\ in Fig.~\ref{qcdgap} but
in the numerator on the l.h.s. This makes the equation nonlinear and hence its
self-consistent solution can exhibit properties inaccessible in perturbation
theory.  Using simple models of the equation's kernel one can unambiguously
establish that $B(p^2)\neq 0$ is the favoured solution in the chiral limit if,
and only if, the integrand provides sufficient support on the domain
$k^2\in[0,2]\,$GeV$^2$; i.e., if the effective coupling in the infrared is
\textit{strong enough}~\cite{cdranu,cdresi}.  A realistic, one-parameter model
of the effective interaction~\cite{mr97} yields the mass function depicted in
Fig.~\ref{quarkM}.  It is important to remember that the existence of a nonzero
solution in the chiral limit is a purely nonperturbative effect. Hence the
domain on which the chiral limit solution and the $u$-quark solution are nearly
indistinguishable is that on which nonperturbative dynamics is dominant in QCD:
where they separate marks the beginning of the perturbative domain and this
point is characterised by a length-scale of $\sim 0.15\,$fm.

The quark condensate is a fundamental fitting parameter in chiral perturbation
theory. In DSE studies it is a calculated quantity that can simply be read-off
from the large-$p^2$ behaviour of the dressed-quark mass function. In Landau
gauge
\begin{equation}
M(p^2) \stackrel{{\rm large}-p^2}{=}\,
\frac{2\pi^2\gamma_m}{3}\,\frac{\left(-\,\langle \bar q q \rangle^0\right)}
           {p^2
        \left(\mbox{$\frac{1}{2}$}\ln\left[\frac{p^2}{\Lambda_{\rm
                QCD}^2}\right]
       \right)^{1-\gamma_m}}\,,
\end{equation}
where $\gamma_m= 12/(33-2 N_f)$, $N_f=4$, is the anomalous mass dimension, and
$\langle \bar q q \rangle^0$ is the gauge-invariant and
renormalisation-point-independent vacuum quark condensate, which is easily
evolved to the mass-scale $\sim 1\,$GeV relevant to chiral perturbation theory.
The one-parameter model of Ref.~\cite{mr97} yields
\begin{equation}
\label{cndstvalue}
 -\langle \bar q q \rangle^0_{1\,{\rm GeV}} = (0.241\,{\rm
GeV})^3,
\end{equation}
a value consistent with recent lattice simulations~\cite{peteradelaide}.  The
condensate's value is a QCD analogue of the Cooper pair density in a BCS
superconductor, and the value in Eq.~(\ref{cndstvalue}) corresponds to a
density of $1.8\,$fm$^{-3}$.  Each sphere in a close-packed sea with this
density would have a radius $r_{\langle \bar q q\rangle}=0.76\,$fm, which is
just $15$\% larger than the pion's charge radius and $13$\% less that the
proton's.  This comparison underscores the importance of a proper description
of DCSB in any attempt to explain low-energy phenomena.

%--------------------------------------------------------------------------------
%\begin{figure}[t]
%\centerline{\epsfig{file=Mp2bonn.eps,width=0.55\textwidth,silent=,clip=}}
%\caption{\label{quarkM} Quark mass function: $M(p^2)=B(p^2)/A(p^2)$, in the
%chiral limit, and for the $u$-quark (isospin symmetry is assumed):
%$m_u^{1\,{\rm GeV}}=5.5\,$MeV, and the $s$-quark: $m_s^{1\,{\rm
%GeV}}=130\,$MeV.  The intersection of the line $p^2=M^2(p^2)$ with a given
%curve gives the Euclidean constituent-quark mass.}
%\end{figure}
%--------------------------------------------------------------------------------
%--------------------------------------------------------------------------------

\section{Mesons}
An aim of contemporary experiments is to explore the transition between the
nonperturbative and perturbative domains in QCD. That requires momentum
transfers at which a theoretical description of the reactions must be
Poincar\'e covariant.  For mesons this means employing an homogeneous
Bethe-Salpeter equation (BSE) to calculate the mass, and the amplitude that
will be used in evaluating necessary matrix elements.  (While QCD's gap
equation is a DSE for the dressed-quark $2$-point function, a propagator, the
BSE is a DSE for a $3$-point function; i.e., a vertex.)  There is a direct
connection between the Bethe-Salpeter bound state equation and the vertices
that appear in the Ward-Takahashi identities that are a true field theoretical
representation of current conservation. The simplest of these identities relate
the vertices to the dressed-quark propagator and therefore their fulfillment is
only possible if there is a tight connection between the kernel in QCD's gap
equation and that in the Bethe-Salpeter equations.

It is a feature of quantum field theory that the Bethe-Salpeter kernel cannot
be written in a closed form and hence all concrete calculations must employ a
truncation.  This is also true of the gap equation's kernel. It is commonplace
to find calculations that ignore the constraints applied by the Ward-Takahashi
identities, and employ kernels in the the gap equation and the BSE that are
incompatible.  Such studies necessarily violate the chiral symmetry constraints
which the success of chiral perturbation theory has shown to be so important in
low-energy QCD.  It need not be thus.

\subsection{A Mass Formula}
There is at least one practical, systematic, symmetry preserving truncation of
the DSEs~\cite{truncscheme}.  It has been used~\cite{mrt98} to prove
Goldstone's theorem in QCD, to obtain quark-level Goldberger-Treiman relations
that relate the scalar functions in the pion's Bethe-Salpeter amplitude to
those in the dressed-quark propagator, and to derive a mass-formula for flavour
nonsinglet pseudoscalar mesons:
\begin{equation}
\label{massform}
f_H^2 \, m_H^2 = - \langle \bar q q \rangle^H_\zeta \, {\cal M}^H_\zeta .
\end{equation}
In this equation: ${\cal M}^H_\zeta = m_\zeta^{q_1} + m_\zeta^{q_2}$ is the sum
of the current-quark masses of the meson's constituents, with $\zeta$ the
renormalisation point; the electroweak decay constant is obtained from
\begin{equation}
\label{fH} f_H \, P_\mu = Z_2(\zeta,\Lambda)
\int^\Lambda \! \frac{d^4 q}{(2\pi)^4} \,
{\rm tr}\left[\left( \mbox{\small $\frac{1}{2}$} T^H\right)^{\rm t} \gamma_5
\gamma_\mu {\cal S}(q_+) \Gamma^H(q;P) {\cal S}(q_-)\right],
\end{equation}
with $Z_2(\zeta,\Lambda)$ the quark wavefunction renormalisation, $\Lambda$ the
mass-scale in a translationally invariant regularisation of the integral,
$q_\pm = q\pm P/2$ ($P_\mu$ is the meson's momentum and $\Gamma^H$ is its bound
state amplitude), and, e.g., $T^{\pi^+}= (\lambda^1 + i \lambda^2)/2$, where
$\{\lambda^j,j=1,\ldots, 8\}$ are the Gell-Mann matrices, and $(\cdot)^{\rm t}$
denotes matrix transpose; and the in-hadron condensate is
\begin{equation}
i \langle \bar q q \rangle^H_\zeta  = f_H \, Z_4(\zeta,\Lambda)
\int^\Lambda \!  \frac{d^4 q}{(2\pi)^4} \,
{\rm tr}\left[\left( \mbox{\small $\frac{1}{2}$} T^H\right)^{\rm t} \gamma_5
{\cal S}(q_+) \Gamma^H(q;P) {\cal S}(q_-)\right].
\end{equation}
The $Z_2$ on the r.h.s.\ of Eq.~(\ref{fH}) ensures that $f_H$ is gauge
invariant, and cutoff and renormalisation-point independent; i.e., that it is
observable.  (Equation.~(\ref{fH}) is the field theoretical expression for the
pseudovector projection of the pion's wavefunction at the origin in
configuration space.)  Furthermore, the quark-level Goldberger-Treiman
relations and Eq.~(\ref{fH}) make plain that in the presence of DCSB the
magnitude of the pion's leptonic decay constant is set by the constituent-quark
mass.  The same is true for $\langle \bar q q \rangle^H_\zeta$, which, in the
chiral limit, is identical to the vacuum quark condensate~\cite{mrt98}.  The
factor $Z_4$ ensures that the in-hadron condensate is gauge and cutoff
independent, and that its renormalisation-point dependence is precisely that
required to ensure $\langle \bar q q \rangle^H_\zeta {\cal M}_\zeta$ is
renormalisation-point {\it in}-dependent.

Equation~(\ref{massform}) is written suggestively: it has the appearance of the
Gell-Mann--Oakes--Renner relation, and it can be shown~\cite{mrt98} that for
small current-quark masses it does indeed coincide with that formula.  The new
aspect of the equation is that it is valid \textit{independent} of the
current-quark mass of the constituents: the DSE derivation assumes nothing
about the size of $m_\zeta^{q_1,q_2}$.  It has consequently been used to prove
that the mass of a heavy pseudoscalar meson rises linearly with the mass of its
heaviest constituent~\cite{mishaheavy}.  Equation~(\ref{massform}) is a single
mass formula that unifies the light- and heavy-quark domains.  It also
provides~\cite{qciv} an understanding of recent lattice-QCD data on the
current-quark mass-dependence of pseudoscalar meson masses.

The truncation scheme of Ref.~\cite{truncscheme} is the foundation for a
phenomenological model that has been used to very good effect in describing
light-quark mesons~\cite{mr97,pieterhere}.  That model is the only one to
predict~\cite{pieterpion} a behaviour for the pion's electromagnetic form
factor that agrees with the results of a recent Hall~C
experiment~\cite{volmer}.  The large-$Q^2$ behaviour of the form factor can be
obtained algebraically and one finds~\cite{mrpion} $Q^2 F_\pi(Q^2)=\,$const.,
up to logarithmic corrections, in agreement with the perturbative-QCD
expectation. This result relies on the presence of pseudovector components in
the pion's Bethe-Salpeter amplitude, which is guaranteed by the quark-level
Goldberger-Treiman relations proved in Ref.~\cite{mrt98}.

\subsection{Pion's Valence Quark Distribution Function}
\begin{figure}[t]
\parbox{.55\textwidth}{\epsfig{file=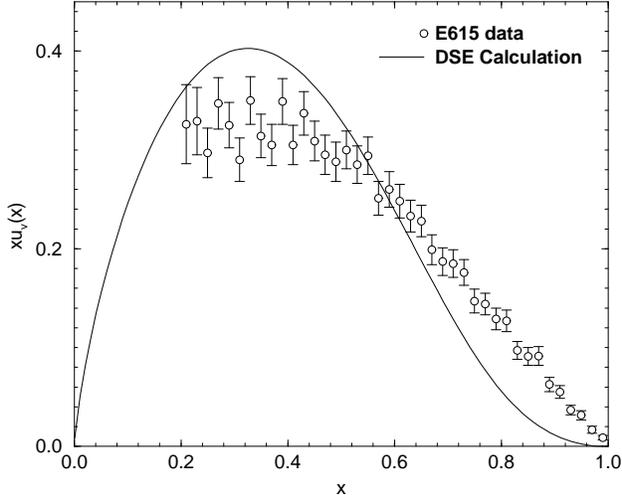,width=.5\textwidth,silent=,clip=}}
\hfill
\parbox{.45\textwidth}{\caption{\label{uvxplot} DSE result for
$x \,u^\pi_V(x)$~\protect\cite{uvxdse}, evolved to $\mu^2=16\,$GeV$^2$ using
the first-order, nonsinglet renormalisation group equation, for direct
comparison with the Drell-Yan data~\protect\cite{DY}.}}
\end{figure}

The DSEs provide a chiral-symmetry preserving, dynamical approach to QCD, which
easily captures the dichotomous nature of the pion as: 1) QCD's Goldstone mode;
and 2) a bound state of quarks with large constituent-masses, and unifies the
low- and high-$Q^2$ domains.  They are therefore well-suited to a study of the
pion's valence-quark distribution function, $u^\pi_V(x)$.  (This is a
measurable expression of the pion's quark-gluon substructure but it cannot be
calculated in perturbation theory.)  Since $\pi$ targets are scarce other means
must be employed to measure $u^\pi_V(x)$ and one approach is to infer it from
$\pi N$ Drell-Yan~\cite{DY}, which yields $u^\pi_V(x) \propto (1-x)$ for
$x\simeq 1$. However, a DSE calculation predicts~\cite{uvxdse}:
\begin{equation}
u^\pi_V(x) \stackrel{x\sim 1}{\propto} (1-x)^2\,,
\end{equation}
at a resolving scale of $\mu^2 \simeq 1\,$GeV$^2$, and while this conflicts
with the extant experimental result, as apparent in Fig.~\ref{uvxplot}, it is
consistent with the perturbative-QCD expectation~\cite{uvxpQCD}.  The
discrepancy is very disturbing because a verification of the experimental
result would present a profound threat to QCD, even challenging the assumed
vector-exchange nature of the interaction.  The DSE study~\cite{uvxdse} has
refocused attention on this disagreement, and is the catalyst for a resurgence
of interest in $u^\pi_V(x)$ and proposals for its remeasurement~\cite{roy}.

%--------------------------------------------------------------------------------
%--------------------------------------------------------------------------------

\section{Baryons}
A direct analogy to treating mesons via the Bethe-Salpeter equation is to
describe baryons using a Poincar\'e covariant Fadde'ev equation.  Of course,
that equation also involves a kernel about which assumptions must be made in
order to arrive at a tractable problem.  Here the truncation scheme of
Ref.~\cite{truncscheme} provides, {\it a posteriori}, a basis for treating
baryons as quark-diquark composites using a Fadde'ev equation of the type
proposed in Ref.~\cite{regfe}.  Therein two quarks are always correlated as a
colour-antitriplet diquark quasiparticle (because ladder-like gluon exchange is
attractive in the $\bar 3_c$ quark-quark scattering channel) and binding in the
nucleon is effected by the iterated exchange of roles between the dormant and
diquark-participant quarks.  A first numerical study of this Fadde'ev equation
was reported in Ref.~\cite{cjbfe}, and following that there have been numerous
more extensive analyses and applications; e.g., Refs.~\cite{bentz,oettel},
which are reviewed in Ref.~\cite{mikehere}.

\subsection{Fadde'ev Amplitude {\it Ansatz}}
In developing an efficacious phenomenology it is possible to bypass solving the
Fadde'ev equation, which can be a numerically intensive process, and employ a
product {\it Ansatz} for the nucleon's bound state amplitude; an approach
kindred to that which is still employed fruitfully in the study of meson
properties, e.g., Ref.~\cite{mikesplitting}.  The simplest {\it Ansatz} retains
only a scalar ($0^+$) diquark correlation and models the nucleon's amplitude
as~\cite{edm}:
\begin{equation}
\psi(p_i,\alpha_i,\tau_i) \propto \varepsilon_{c_1 c_2 c_3} \, [\Gamma^{0^+}(
\mbox{\small$\frac{1}{2}$}(p_1-p_2);K) ]_{\alpha_1 \alpha_2}^{\tau_1 \tau_2} \;
\Delta^{0^+}(K) \; [{\cal S}(\ell;P) u(P)]_{\alpha_3}^{\tau_3}\,,
\end{equation}
where: $(p_i,\alpha_i,\tau_i)$ are the momenta, spin and isospin labels of the
quarks comprising the nucleon; $\varepsilon_{c_1 c_2 c_3}$ is the Levi-Civit\`a
symbol that gives the colour singlet factor; $P=p_1+p_2+p_3$, $K=p_1+p_2$ and
$\ell= (-p_1-p_2 + 2 p_3)/3$; $\Delta^{0^+}(K)$ is a pseudoparticle propagator
for the scalar diquark formed from quarks $1$ and $2$, and $\Gamma^{0^+}\!$ is
a Bethe-Salpeter-like amplitude describing their relative momentum correlation;
${\cal S}$, a $4\times 4$ Dirac matrix, describes the relative quark-diquark
momentum correlation; and $u(P)$ is a free-nucleon spinor.  The unknown
functions are parametrised as (${\cal F}(x)=(1-{\rm e}^{-x})/x$):
\begin{eqnarray}
\Delta^{0^+}(K) & = & \frac{1}{m_{0^+}^2}\,{\cal F}(K^2/\omega_{0^+}^2)\,,
\;
\Gamma^{0^+}(k;K) = \frac{1}{{\cal N}^{0^+}} \, C i\gamma_5\, i\tau_2\, {\cal
F}(k^2/\omega_{0^+}^2) \,,\\
{\cal S}(\ell;P) & = & \frac{1}{{\cal N}^{\psi}} \, {\cal
F}(\ell^2/\omega_\psi^2) \left[ I_{\rm D} - \frac{\mbox{\sc
r}}{M}\left(i\gamma\cdot \ell - \ell \cdot \hat P\, I_{\rm D}\right) \right]
\,,
\end{eqnarray}
where: $C=\gamma_2\gamma_4$ is the charge conjugation matrix; ${\cal N}^{0^+}$
is a calculated normalisation factor that guarantees an electric charge of
$1/3$ for the scalar diquark; ${\cal N}^{\psi}$ is a similar factor that
ensures the proton has unit charge; and $\mbox{\sc r}= 0.5$ measures the ratio
of lower to upper component in the nucleon's spinor in the rest frame, which is
taken from Fadde'ev equation studies~\cite{qciv}. The model has three
parameters: $m_0^+$, the diquark's mass; and $\omega_{0^+}$ and $\omega_\psi$.
They have physical intepretations: $d_{0^+} = 1/m_{0^+}$ is the distance over
which a scalar diquark correlation can propagate inside the nucleon; $l_{0^+} =
1/\omega_{0^+}$ is a measure of the mean separation between the quarks in the
scalar diquark; and $l_{\psi} = 1/\omega_{\psi}$ measures the mean separation
between the dormant quark and the diquark.

In the many applications of this {\it Ansatz} the parameters are typically
determined by requiring a least-squares fit to the proton's electromagnetic
form factor, as described in Ref.~\cite{edm}.  The procedure yields (in GeV or
fm, as appropriate)
\begin{equation}
\label{Sets}
\begin{array}{ccc|ccc}
 m_{0^+} & \omega_{0^+} & \omega_\psi &
1/m_{0^+} & 1/\omega_{0^+} & 1/\omega_\psi\\\hline
 0.62 & 0.79 & 0.23 & 0.32 & 0.25 & 0.86
\end{array}
\end{equation}
and it is plain that these values provide an internally consistent picture:
$l_{0^+} < l_\psi$, which means that the dormant quark is spatially separated
from the diquark; and $d_{0^+}< r_p$, the proton's charge radius, so that the
diquark doesn't propagate outside the nucleon.  The subsequent calculated
results are then predictions and tests of the model's fidelity, which can be
gauged from Table~\ref{table1}.

\begin{table}[t]
\parbox{.55\textwidth}{%
\begin{tabular*}
{\hsize}
{l@{\extracolsep{0ptplus1fil}}c@{\extracolsep{0ptplus1fil}}c@{\extracolsep{0ptplus1fil}}c}
%{l|c|c|c}
                  & Obs.         & Calc.    \\ \hline
$(r_p)^2\,($fm$^2$) & ~$(0.87)^2$ & ~$(0.78)^2$ \\
$(r_n)^2\,($fm$^2$) & $-(0.34)^2$& $-(0.40)^2$  \\
$\mu_p \, (\mu_N)$  & $2.79$     & $2.85$   \\
$\mu_n \, (\mu_N)$  &$-1.91$     &$-1.61$    \\
$\mu_n/\mu_p$       &$-0.68$     &$-0.57$    \\
$g_{\pi NN}$        &$13.4$      &$13.9$    \\
$\langle r_{\pi NN}^2\rangle\,($fm$^2$)
                    & $(0.93$-$1.06)^2$  &$(0.63)^2$   \\
$g_A$               &$1.26$    &$0.98$    \\
$\langle r_{A}^2\rangle\,($fm$^2$)
                    &$(0.68\pm 0.12)^2$    &$(0.83)^2$   \\
$g_{\rho NN}$       &$6.4$    &$5.61$  \\
$g_{\omega NN}$     &$7$--$10.5$ & $10.0$ \\\hline
\end{tabular*}}
\hfill
\parbox{.45\textwidth}{\caption{\label{table1} Calculated values of a range of
physical observables obtained~\protect\cite{edm} using the Fadde'ev {\it
Ansatz} parameters in Eq.~(\protect\ref{Sets}).  The ``Obs.''\ column reports
experimental values~\cite{expt} or values employed in a typical meson exchange
model~\protect\cite{harry}.}}
\end{table}

\subsection{Meson Photoproduction}
The same {\it Ansatz} is being used to study photoproduction of mesons from the
nucleon.  These processes are important for developing an understanding of the
structure of nucleon resonances and in searching for ``missing'' resonances;
i.e., those states predicted by constituent quark models that are hitherto
unobserved.  The aim of these calculations is to provide and constrain the
input to meson exchange models, which are the tool that makes possible a
comparison with data by subsequently incorporating details of the reaction
mechanism.  First results for $\omega$ photoproduction were presented in
Ref.~\cite{hechtadelaide} and, while only the $t$-channel $\pi$-exchange
mechanism was described, the results at forward angles, where this process must
dominate, demonstrate the approach's promise. A calculation of the
contributions to the cross-section from the $s$- and $u$-channel processes is
almost complete.

\begin{figure}[t]
\vspace*{-4ex}
\parbox{.55\textwidth}{\begin{center}
\begin{picture}(144,120)(-72,-72)
\setlength{\unitlength}{0.75pt}
\thicklines
\put (-72,-72){\line(2,1){72}} \put (-66,-60){$N$}
\put (0,-36){\line(2,-1){72}} \put (66,-60){$N$}
\put (0,-36){\circle*{12}}
\multiput (0,-36)(0,9){6}{\line(0,1){6}} \put (9,-9){$\pi^\ast$}
\put (0,18){\circle*{12}}
\multiput (0,18)(20,10){4}{\line(2,1){14}}\put (66,42){$\pi$}
%
%\put (0,18){\line(-2,1){72}}
%
\multiput (0,18)(-20,10){4}{\line(-2,1){14}} \put(-72,42){$\gamma$}
\end{picture}\end{center}}\hfill
\parbox{.45\textwidth}{\caption{\label{photopi} $t$-channel $\pi$-exchange
contribution to the $\pi$-photoproduction amplitude.  In meson exchange models
the $\gamma\pi^\ast\pi$ vertex is usually considered
momentum-independent.}}\vspace*{-4ex}
\end{figure}
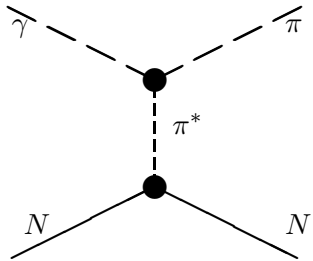

\begin{figure}[t]
\parbox{.55\textwidth}{\epsfig{file=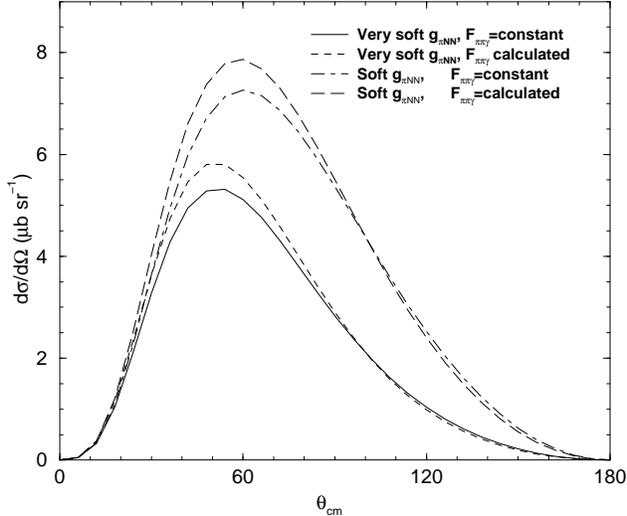,width=.5\textwidth,silent=,clip=}}
\hfill
\parbox{.45\textwidth}{\caption{\label{diffphotopi} $\gamma N \to \pi N$
differential cross-section at $E_\gamma=0.34\,$GeV as obtained solely from the
Born diagram in Fig.~\protect\ref{photopi}.  The solid line is the Born
cross-section obtained~\cite{harryprivate} using the meson exchange model of
Ref.~\protect\cite{harryphotopi} whose parameters were fitted to data in a full
$T$-matrix calculation.  The remaining curves, discussed in the text, are
calculated using DSE input.}}
\end{figure}

Meanwhile the analogous contribution to pion photoproduction, illustrated in
Fig.~\ref{photopi}, has been calculated and the parameter-free results are
depicted in Fig.~\ref{diffphotopi}.  The solid curve is the Born approximation
calculated using an efficacious meson exchange model~\cite{harryphotopi}.  That
model neglects the necessary momentum dependence of the $\gamma\pi^\ast\pi$
vertex, for which the DSEs make a prediction.  Including the calculated vertex
yields the short-dashed curve in Fig.~\ref{diffphotopi}, which is not
materially different.  \textit{A posteriori}, this justifies the meson exchange
model expedient of neglecting quark-gluon substructure for these vertices.  Of
course, with the new DSE predictions, quantitative improvements are now
possible.  The dashed and dot-dashed curves in Fig.~\ref{diffphotopi} were
obtained using the calculated dipole width for the $\pi N N$ vertex: $F_{\pi N
N}(t_\pi)$, which is larger ($\lesssim$ 2) than that fitted to data in
Ref.~\cite{harryphotopi}. The effect is significant in this case.  However,
there are preliminary indications that the particularly soft form factor used
in the meson exchange model is implicitly also accounting for
off-nucleon-mass-shell suppression in the form factor.  It is a known property
(see, e.g., Ref.~\cite{mikesplitting}) that vertices describing the interaction
of three composite objects provide suppression when any one of the attached
legs is off-shell.  The DSE prediction for the strength of this effect will
facilitate the qualitative improvement of meson exchange models by making
possible the explicit representation of this phenomenon.

\subsection{Neutron's Electric Dipole Moment}
A well-founded description of the nucleon makes many applications possible,
even constraining extensions of the Standard Model, which is an important focus
of contemporary nuclear and particle physics.  It has long been known that the
possession of an electric dipole moment (EDM) by a spin-$1/2$ particle would
signal the violation of time-reversal invariance.  Any such effect is likely
small, given the observed magnitude of $CP$ and $T$ violation in the neutral
kaon system, and this makes neutral particles the obvious subject for
experiments: the existence of an electric monopole charge would overwhelm most
signals of the dipole strength.  It is therefore natural to focus on the
neutron, which is the simplest spin-$\frac{1}{2}$ neutral system in nature.
Attempts to measure the neutron's EDM, $d_n$, have a long history and
currently~\cite{edmexpt}
\begin{equation}
\label{dnub} |d_n| < 6.3 \times 10^{-26}\,e\,{\rm cm}\; (90\%\,{\rm C.L.}).
\end{equation}
(NB.\ $e/(2 M_n) = 1.0 \times 10^{-14}\,e\,{\rm cm}$.  Therefore, writing $d_n
= e h_n/(2 M)$, where $M$ is the neutron's mass and $h_n$ is its ``gyroelectric
ratio,'' Eq.~(\ref{dnub}) corresponds to $|h_n| < 6.0 \times 10^{-12}$.)

This experimental constraint on $d_n$ has been very effective in ruling out
candidates for theories that enlarge the Standard Model.  That is true because
in the Standard Model the first nonzero contribution to a free quark's EDM
appears at third order and involves a gluon radiative correction (i.e.,
O$(\alpha_s\,G_F^2)$, for the same reason that flavour-changing neutral
currents are suppressed: the GIM mechanism) so that~\cite{edm}
\begin{equation}
\label{ddSM} d_n^{\rm SM} \lesssim 10^{-34}\,e\,{\rm cm}\,.
\end{equation}
This is seven orders-of-magnitude less than the experimental upper bound.
However, the Standard Model is peculiar in this regard and candidates for its
extension typically contain many more possibilities for $CP$ and $T$ violation,
which are not {\it a priori} constrained to be small.  Hence Eq.~(\ref{dnub})
is an important and direct constraint on these extensions because
Eq.~(\ref{ddSM}) indicates that the Standard Model contribution to $d_n$ cannot
interfere at a level that could currently cause confusion.

%--------------------------------------------------------------------------------
%--------------------------------------------------------------------------------

Extensions of the Standard Model are typically used to predict current-quark
EDMs, and to proceed from these to a result for $d_n$ one must have an
understanding of the relation between current-quarks and constituent-quarks,
and a reliable model of the neutron. The DSEs give both: the first from the gap
equation, and the second from the Fadde'ev equation studies and their
phenomenological application. Putting these elements together one
arrives~\cite{edm} at a quantitative relation between the current-quark's
gyroelectric ratio and that of the neutron: $h_n \approx - 91 \,h_d$.  Here the
very large magnifying factor: $91$, owes its appearance to DCSB, which turns
the current-quark into the constituent-quark, and it means that
Eq.~(\ref{dnub}) applies the following bound:
\begin{equation}
\label{hdU} |h_d| < 7.4 \times 10^{-14}\,.
\end{equation}
(NB.\ Most extensions of the Standard Model predict $h_u\ll h_d$.)  This
DCSB-tightened bound very much threatens the viability of a popular
three-Higgs-boson model of spontaneous $CP$ violation~\cite{weinberg}.
Furthermore, and importantly, the result in Eq.~(\ref{hdU}) is independent of
the model used to calculate $h_d$ and hence can be applied directly to
constrain any extension of the Standard Model.

\section{Epilogue}
Owing to dynamical chiral symmetry breaking in QCD the pion appears as both a
Goldstone mode and a bound state of massive constituents.  The existence of at
least one systematic, symmetry preserving truncation scheme makes a detailed
understanding and explanation of this phenomenon possible using the
Dyson-Schwinger equations (DSEs). This is the starting point for a successful
phenomenology of strong interaction phenomena, which quantitatively and
qualitatively unifies the low- and high-energy regimes.  Since the DSEs
maintain contact with perturbation theory, the model-dependence is restricted
to a statement about the infrared behaviour of the quark-quark interaction;
i.e., the unknown nature of the long-range force in QCD.  This remaining
model-dependence is a virtue because it makes possible the correlation of
observables via a parametrisation of this infrared behaviour and hence the use
of experiments as a probe of the confining force.

%%%%%%%%%%%%%%%%%%%%%%%%%%%%%%%%%%%%%%%%%%%%%%%%%%%%%%%%%%%%%%%%%%%%%%%%%%%%%%%%%%
\acknowledgments{This work was supported by the US Department of Energy,
Nuclear Physics Division, under contract no.~\mbox{W-31-109-ENG-38}.}

%%%%%%%%%%%%%%%%%%%%%%%%%%%%%%%%%%%%%%%%%%%%%%%%%%%%%%%%%%%%%%%%%%%%%%%%%%%%%%%%%%
%____________________________________________________________
%  Start references here:

\end{document}